\documentclass[12pt,a4paper]{iopart}

\usepackage{graphicx}

\begin{document}

\title[Magnetic phase evolution in the spinel compounds 
Zn$_{1-x}$Co$_x$Cr$_2$O$_4$]
{Magnetic phase evolution in the spinel compounds 
Zn$_{1-x}$Co$_x$Cr$_2$O$_4$}

\author{Brent C. Melot, Jennifer E. Drewes, and Ram Seshadri}
\address{Materials Department and Department of Chemistry and Biochemistry\\
University of California, Santa Barbara CA 93106} 
\eads{bmelot@mrl.ucsb.edu, jennidrewes@umail.ucsb.edu, seshadri@mrl.ucsb.edu}

\author{E. M. Stoudenmire}
\address{Department of Physics\\
University of California, Santa Barbara CA 93106}
\eads{miles@physics.ucsb.edu}

\author{Arthur P. Ramirez}
\address{LGS, 15 Vreeland Road, Florham Park, New Jersey NJ 07932} 
\eads{apr@LGSInnovations.com}

\begin{abstract}

We present the magnetic properties of complete solid solutions of 
ZnCr$_2$O$_4$ and CoCr$_2$O$_4$: two well-studied oxide spinels with very 
different magnetic ground states. ZnCr$_2$O$_4$, with non-magnetic $d^{10}$ 
cations occupying the A site and magnetic $d^3$ cations on the B site, is a 
highly frustrated antiferromagnet. CoCr$_2$O$_4$, with magnetic $d^7$ cations 
(three unpaired electrons) on the A site as well, exhibits both N\'eel 
ferrimagnetism as well as commensurate and incommensurate non-collinear 
magnetic order. More recently, CoCr$_2$O$_4$ has been studied extensively for 
its polar behavior which arises from conical magnetic ordering. Gradually 
introducing magnetism on the A site of ZnCr$_2$O$_4$ results in a transition 
from frustrated antiferromagnetism to glassy magnetism at low concentrations of 
Co, and eventually to ferrimagnetic and conical ground states at higher 
concentrations. Real-space Monte-Carlo simulations of the magnetic 
susceptibility suggest that the first magnetic ordering transition and
features of the susceptibility across $x$ are captured by near-neighbor
self- and cross-couplings between the magnetic A and B atoms.
We present as a part of this study, a method for displaying 
the temperature dependence of magnetic susceptibility in a manner which helps 
distinguish between compounds possessing purely antiferromagnetic interactions 
from compounds where other kinds of ordering are present.  

\end{abstract}

\maketitle

\section{Introduction}

Geometrically frustrated spin systems\cite{Ramirez1994} have been extensively 
investigated because of the fascinating fundamental physics they display. 
More recently, they have attracted interest due to their frequently possessing
ground states with spiral magnetic structures.\cite{KimuraAnnuRev2007}
Recent interest in such structures has been engendered by the rediscovery
of systems where the magnetic ordering does not have a center of 
symmetry, and there is perforce a loss of a center of symmetry in the crystal 
structure as well. This helps to couple lattice and spin degrees of freedom 
and gives rise to a plethora of magnetoelectric 
phenomena.\cite{Kimura2003,Katsura2005,Kenzelmann2005,Mostovoy2006,
Sergienko2006,Yamasaki2006,Lawes2006,Cheong2007,Tackett2007} 

\begin{figure}
\centering \includegraphics[width=8cm]{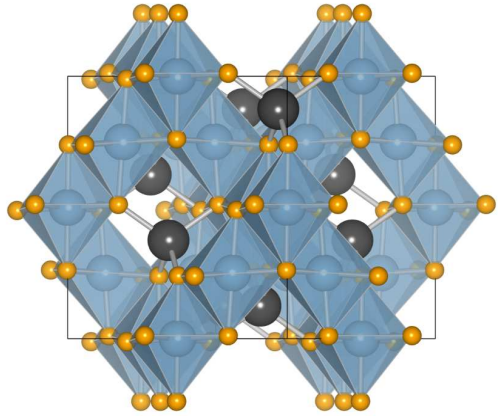}
\caption{(Color online) Spinel AB$_2$O$_4$ structure showing edge-shared laths 
of BO$_6$ octahedra (blue-grey) with B--B magnetic coupling across the edges.
Tetrahedrally coordinated A atoms (dark grey) connect the octahedral laths,
and each A atoms has four B near-neighbors. Oxygen are orange.
\label{fig:struc}}
\end{figure}

The spinel structure type (displayed in figure\,\ref{fig:struc}) with the 
general formula AB$_2$X$_4$, has been extensively studied in this light 
because of its ability to host magnetic cations on both its tetrahedrally 
coordinated A sublattice as well as its octahedrally coordinated B sublattice. 
Additionally, both sublattices are geometrically frustrated with octahedral B 
sites forming a pyrochlore network, and the tetrahedral sites forming a diamond 
lattice. The former intrinsically displays geometric magnetic frustration
associated with the difficulty of decorating the vertices of a tetrahedra 
with spins that are antialigned. The latter is frustrated because of competing 
near- and next-near neighbor 
interactions.\cite{Fritsch2004,Buettgen2004,Bergman2007}

Conventional wisdom, however, is that the 
dominant magnetic interactions are between the A and the B 
sublattices provided there are magnetic ions on both sites. Adding to the
complex nature of magnetic interactions in the spinel structure is that 
superexchange interactions between A-sites are mediated by atoms 
on the B-site (even when the B site is non-magnetic), and in the same vein,
B-B next-nearest-neighbor interactions are mediated by the A site 
ion.\cite{Blasse1963} In the cubic spinel, the number of near neighbors (NN) 
and next-near-neighbors (NNN) for A are 4A(NN), 12B(NN), 12A(NNN), and 
16B(NNN). For B, the distribution is 6A(NN), 6B(NN), 8A(NNN), and 12B(NNN). 
In addition, most of the coupling pathways are multiply degenerate. These 
result in next-near neighbors contributing in an important manner to the 
overall magnetic ordering behavior.\cite{Fritsch2004,Buettgen2004} 
The high connectivity of the spinel lattice 
has been shown to give rise to a vast array of interesting magnetic
phenomenon.
\cite{Lawes2006,Yamasaki2006,Tackett2007,Baltzer1965,Baltzer1966,Ramirez1997,Rudolf2007,Ortega-San-Martin2008,Tokura2008} 

Chromium spinels, where the B sublattice is fully occupied by
Cr$^{3+}$, present a unique opportunity to examine these complex magnetic
interactions due to the fact that Cr$^{3+}$ will not invert to the
tetrahedral A sites unlike many other transition metals, as a result of the 
very strong crystal field stabilization of octahedral $d^3$ cation 
($t_{2g}^3$ crystal field). Here we  prepare and study clean samples of solid 
solutions of ZnCr$_2$O$_4$ and CoCr$_2$O$_4$, two well studied chromium 
spinels with very 
different magnetic ground states. ZnCr$_2$O$_4$, with a non-magnetic cation 
occupying the A site, undergoes a spin-driven Jahn-Teller-like distortion at
low temperatures which results in a N\'{e}el-type antiferromagnetic
order.\cite{Kino1971,Lee2000} CoCr$_2$O$_4$, with magnetic cations on both
sublattices, displays collinear ferrimagnetic as well as non-collinear
spiral magnetic ordering.\cite{Tomiyasu2004}  By replacing the Zn
with Co our goal is to introduce magnetic interactions into an otherwise
non-magnetic lattice and observe the effect this would have on the magnetic
frustration of the system.  This series is particularly interesting
because it provides the opportunity to study the effects of dilute magnetic
interactions in the absence of changes to the superexchange angle
(Co$^{2+}$ and Zn$^{2+}$ have very similar ionic radii) or site disorder
(because of the strong octahedral site preference of Cr$^{3+}$). We find 
for low concentrations of Co, a seeming increase in frustration resulting 
from spin disorder which results in a glassy magnetic state. At higher Co 
concentrations there is evidence that the spins form non-collinear structures 
reminiscent of the CoCr$_2$O$_4$ end member.  

Monte Carlo simulations of Heisenberg spin systems have been performed which
allow comparisons with the experimental magnetic susceptibility, and yield
the appropriate range of $J$ couplings between the different sites to be 
estimated. They also suggest the minimal models required to describe the 
gross aspects of the magnetic susceptibility in these systems.
We also demonstrate here that appropriate scaling of Curie-Weiss plots of the 
inverse magnetic susceptibility as a function of temperature allows for systems
with purely antiferromagnetic interactions to be distinguished by systems with
more complex interactions. 

\section{Experimental details}

Polycrystalline samples in the series were prepared by solid state routes. 
Appropriate stoichiometric amounts of cobalt oxalate 
(CoC$_2$O$_4$$\cdot$2H$_2$O), ZnO, and Cr$_2$O$_3$ were mixed and ground with
ethanol in an agate mortar.  The powders were then pressed into 13\,mm
pellets and calcined in alumina crucibles at 800$^\circ$C for 12\,h.  
These pellets were then reground, pressed back into pellets, and fired at
1150$^\circ$C for 12\,h.   The pellets were then briefly annealed at
800$^\circ$C for 24\,h. During all heatings, the pellets were placed on a
bed of powder with the same stoichiometry to minimize reaction with the
crucible. X-ray diffraction patterns were obtained using Cu-K$\alpha$ 
radiation on a Philips XPERT MPD diffractometer operated at 45\,kV and 40\,mA.
Phase purity was subsequently determined by refining the patterns using the
Rietveld method as implemented in the {\sc xnd} Rietveld
code.\cite{Berar1998}  DC magnetization measurements were carried out using 
a Quantum Design SQUID magnetometer.

\section{Computational Details}

Classical Monte Carlo simulations of the magnetic behavior of the system were
performed using the ALPS project's \textsc{spinmc} application~\cite{ALPS}.
We  generated a lattice with the appropriate fraction of A sites occupied for 
each simulation run in order to perform disorder averaging. The magnetic 
interactions were modeled by a nearest-neighbor Heisenberg Hamiltonian with
antiferromagnetic couplings $J_{BB}$, $J_{AB}$ and $J_{AA}$ where the
subscripts specify the sublattice types of the two spins connected by a given
interaction. Since the number of $J_{BB}$ interactions in the system does not
vary with $x$, the other two couplings are varied relative to 
$J_{BB}$, which is set to unity (in Monte-Carlo units) throughout. 

The inverse magnetic susceptibility of each simulated system was fit to the
experimental data by scaling both sets of data such that they satisfy the 
Curie-Weiss relationship at at high temperature. This is the same method used 
to plot the experimental data described below -- see equation (\ref{eqn:CW2}).
It should be noted, however, that in these simulations, 
``high temperature'' means high relative to the
temperature at which the inverse susceptibility strongly deviates from linear
behavior. As discussed in what follows, the experiments could not actually be 
performed above $T=|\Theta_{\mbox{\tiny CW}}|$ where the Curie-Weiss law 
is strictly applicable.
Numerical simulations could indeed verify the following analytic expression 
for the Curie-Weiss temperature of the nearest-neighbor model 

\begin{equation}
\Theta_{CW} = \frac{-2 S(S+1)}{3 (1+\frac{x}{2})} (3 J_{BB} + 6x
J_{AB} + x^2 J_{AA}) \label{eqn:analytic_cw} \end{equation}

\noindent After a reasonable fit to the susceptibility was found for 
each experimental dataset, corresponding to the different $x$ values, 
the simulations were run again using a newly generated 
random lattice as many as 3 times to determine the effect of
disorder. A disorder average was then performed by averaging the resulting
inverse susceptibilities and the disorder error bars were taken to be the
standard deviation of this average. Finally, it should be noted that since
the simulations were run until the Monte Carlo error was negligible (relative
errors of about $10^{-4}$), all numerical error bars shown below represent only
the disorder error bars with the exception of the disorder-free $x=1.0$ end
member.  

\section{Results and Discussion} 

\begin{figure}
\centering \includegraphics[width=8cm]{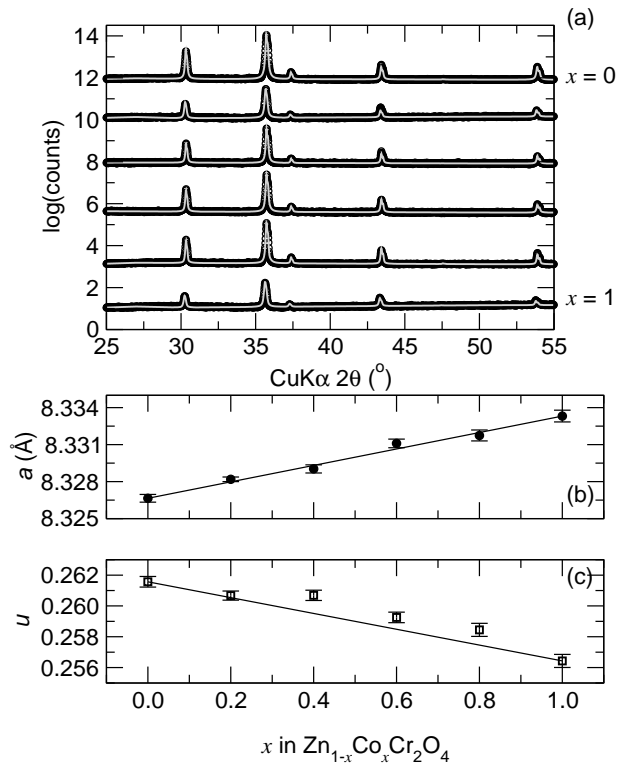} 
\caption{A portion of the room temperature x-ray diffraction patterns of 
Zn$_{1-x}$Co$_x$Cr$_2$O$_4$.
(a) Points are data and solid grey lines represent the Rietveld least-squares
analysis. (b) Evolution of the cubic cell parameter  with Co content $x$. The
line connecting end members is drawn as a guide to the eye to illustrate the
V\'egard law. (c) Shows the steady decrease with $x$  of the internal
positional parameter $u$ of oxygen. \label{fig:xrd}}
\end{figure}

A portion of the X-ray diffraction patterns of the solid solution series 
are presented in figure\,\ref{fig:xrd}. All samples were found to be single 
phase with no extraneous peaks. The patterns were fit using the Rietveld 
method to the normal cubic spinel crystal structure with appropriate mixed 
occupancies on the A site. The cubic cell parameter and the position of the
oxygen atom $(u, u, u)$ in the unit cell were obtained from the refinement.
Because Co and Zn are relatively similar in terms of their x-ray scattering
powers, no attempt was made to extract the relative concentrations of these 
ions in the lattice. 

The cell parameter $a$ and oxygen position $u$ extracted from
the Rietveld refinements are presented in figure\ref{fig:xrd}(b) and
(c).  Co$^{2+}$ and Zn$^{2+}$ in tetrahedral coordination have very similar
ionic radii, 0.58\,\AA\/ and 0.60\,\AA\/ respectively,\cite{Shannon1976}
with Co$^{2+}$ the slightly smaller ion. Increasing Co in the compounds should 
therefore lead to a contraction of the cell parameter. We find, however, a 
small expansion of the unit cell volume. This is consistent with previously 
reported structures of the end members.\cite{Hill1979} A possible reason for 
the expansion is that tetrahedral Co$^{2+}$ is more ionic than Zn$^{2+}$ and the 
substitution of Zn by Co introduces A--B cation-cation repulsions which expand 
the lattice. Expansion in the cell parameters in order to minimize repulsion 
between neighboring cations has been suggested in other spinel 
systems.\cite{Nakatsuka2006}

Changes in the $u$ parameter associated with O atoms in the unit cell occur
as  the structure tries to accommodate cations of different sizes.  When 
$u=0.25$, the anions are in an ideal cubic-close-packed arrangement, with 
perfect CrO$_6$ octahedra, while values of $u>0.25$ indicate an increase in 
size of the tetrahedron with a corresponding shrinkage and trigonal
compression of the octahedron.\cite{Hill1979} As seen from
figure\,\ref{fig:xrd}(c), there  is a systematic decrease in $u$ with $x$
which brings it closer to the ideal  value. This decrease implies that the
tetrahedral sites shrink across the  series, in keeping with the smaller size
of Co$^{2+}$, but in contradiction  with the unit cell expansion. 

\begin{table}
\caption{Results from fitting the inverse magnetic susceptibility data to
the Curie-Weiss equation. $\Theta_{CW}$ is taken as the $x$-intercept of the
inverse susceptibility curve while $T_c$ is chosen as the first point of 
inflection in the first derivative of the susceptibility with respect to 
temperature. $f$ is defined as the absolute ratio of $\Theta_{CW}$ to $T_c$. 
The experimental effective moment is calculated using the relation 
$\mu_{\rm{eff}}^2 = 3Ck_B/N$ where $C$ is the slope of the inverse 
susceptibility versus temperature. The estimated moments (spin-only, 
and unquenched) are obtain using 
$\mu_{\rm{eff}} = \sqrt{ \mu_{\rm{Co}}^2 + 2\mu_{\rm{Cr}}^2}$. 
\label{table:magnetism}}
\begin{center}
\begin{tabular}{lcccccc}
\hline
\hline
\ & \multicolumn{3}{c}{$\mu_{\rm{eff}}$ ($\mu_B$)} & $\Theta_{CW}$ (K)
                                                            & $T_c$ (K) & $f$\\
     \          & measured    & spin-only    & unquenched    &\           & \   & \ \\
\hline
ZnCr$_2$O$_4$   & 5.2  & 5.5  & 5.5    & -298       & 19        & 16 \\
$x$ = 0.2       & 5.7  & 5.8  & 6.0    & -380       & 8         & 51 \\
$x$ = 0.4       & 6.0  & 6.0  & 6.4    & -424       & 15        & 28 \\
$x$ = 0.6       & 6.9  & 6.3  & 6.8    & -592       & 20        & 30 \\
$x$ = 0.8       & 6.7  & 6.5  & 7.2    & -488       & 55        & 9  \\
CoCr$_2$O$_4$   & 7.5  & 6.7  & 7.6    & -568       & 75        & 8  \\
\hline
\hline
\end{tabular}
\end{center}
\end{table}

\begin{figure}
\centering \includegraphics[width=9cm]{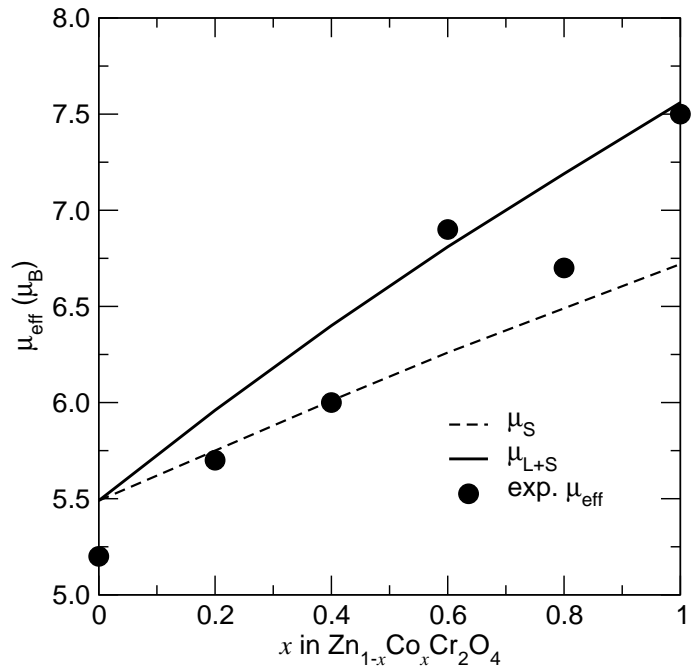}
\caption{Effective magnetic moment obtained from the Curie-Weiss fit to the 
high temperature inverse susceptibility data for Zn$_{1-x}$Co$_x$Cr$_2$O$_4$. 
The two lines represent the expected values calculated for a fully quenched 
($\mu_S$) Co$^{2+}$ and Cr$^{3+}$, and a fully unquenched ($\mu_{L+S}$) 
Co$^{2+}$ orbital contribution to the magnetic moment.
\label{fig:moment}}
\end{figure}

The high temperature (200\,K to 300\,K) susceptibility of the samples was
fit to the Curie-Weiss equation, $\chi = C/(\chi - \Theta_{CW}$), to
obtain the effective paramagnetic moment $\mu_{\rm{eff}}$ from the Curie 
constant $C$, and the Curie-Weiss intercept $\Theta_{CW}$. These values are 
presented in table\,\ref{table:magnetism}. The orbital-quenched, spin-only 
effective moment ($\mu_S = 2\sqrt{S(S+1)}$) for free ions of both Co$^{2+}$ 
and Cr$^{3+}$ is 3.88\,$\mu_B$. Figure \ref{fig:moment} shows an increase 
in the experimental $\mu_{\rm{eff}}$ with increasing values of $x$. It is 
noted that samples with $x\le0.4$ agree well with the spin-only moment. 
$x\ge0.6$ is anomalous in terms of the trend with $x$. The $x$ = 0.8 and 
$x = 1.0$ samples tend increasingly to an effective moment value that is 
closer to the the value expected for the unquenched orbital contribution from
Co$^{2+}$, $\mu_S = \sqrt{4S(S+1) + L(L+1)}$. While this may be unexpected 
for Co$^{2+}$ in a tetrahedral coordination environment which has no orbital 
degeneracy, the presence of the low lying ($e^3t_2^4$) excited state could be 
a source for this increased moment.\cite{Cossee_JPhysChemSolids1960} 
Cr$^{3+}$ is expected to remain strongly orbital quenched. The change from
orbital quenched (spin-only) to unquenched behavior for Co$^{2+}$ in this 
system may have to do with the small changes in structure as $x$ is increased.

It was recently suggested that the introduction of
magnetic cations into ZnCr$_2$O$_4$ has the effect of introducing short
range order in the high temperature paramagnetic regime,\cite{Yan2008} and
this can be a source of the small underestimation observed here. 
In addition, even in dilute spinel systems with Cr$^{3+}$, near-neighbor 
coupling effects are known,\cite{Blazey1966} and such coupling can invalidate 
the assumption made here of independent spins. We note, however, that the 
effective moment we have determined for ZnCr$_2$O$_4$ falls within the range of
previously reported values~\cite{Martinho2001, Sianou2005} which vary from 
5.1-5.6 $\mu_B$ f.u.$^{-1}$.  We were unable to find any previous reports for 
the effective moment of CoCr$_2$O$_4$.

As $x$ increases, so do the magnitudes of the Curie-Weiss $\Theta_{CW}$, 
reflecting the increasing numbers of the dominant antiferromagnetic A--B 
interactions. The increase is monotonic with the exception of $x$ = 0.6,
which seem anomalous in this series. 

\begin{figure*}
\centering \includegraphics[width=14cm]{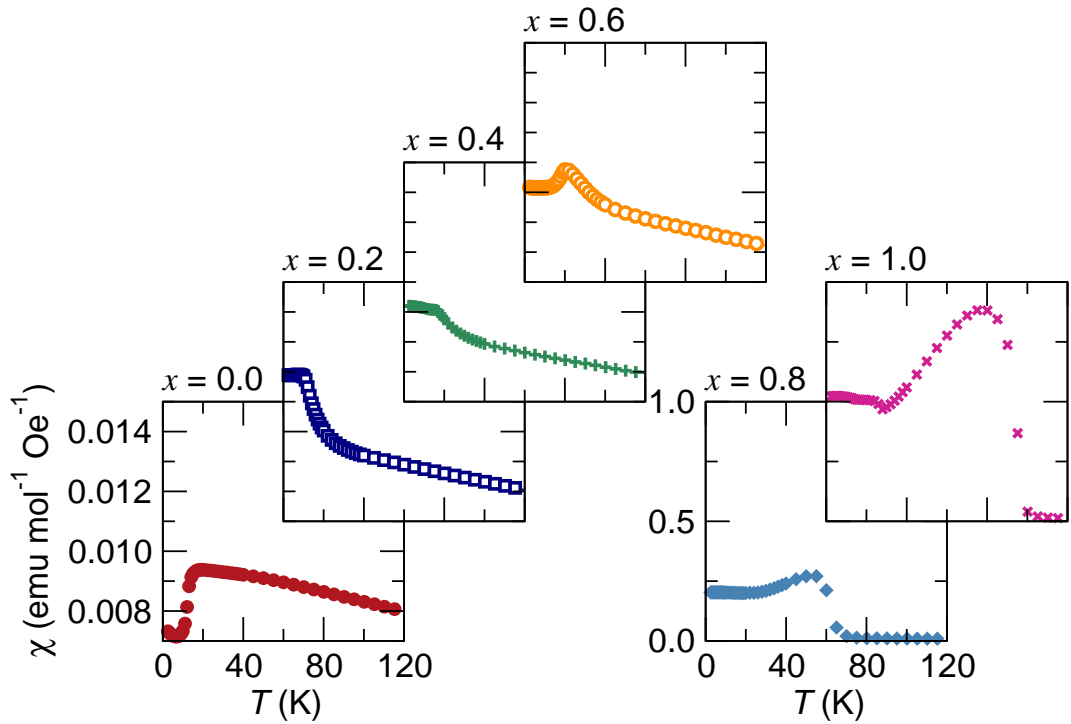}
\caption{(Color online) Field cooling curves were collected under a static DC
field of 1000\,Oe.  Pure ZnCr$_2$O$_4$ sample orders antiferromagnetically
below its N\'{e}el temperature while samples with $0< x \le$ 0.6 exhibit
glassy behavior resulting from the dilute uncompensated antiferromagnetic
interactions because of A--B interactions between Co$^{2+}$ and Cr$^{3+}$.
Samples with $x$ = 0.8 and $x$ = 1.0 order ferrimagnetically below their
highest transition temperature, and then transform to more complex orderings.
\label{fig:fc}}
\end{figure*}

Figure\,\ref{fig:fc} illustrates the temperature dependence of the DC
magnetization with the response of the end members in good agreement with
previously reported data.\cite{Kino1971,Tomiyasu2004}  The sharp
drop in the susceptibility of ZnCr$_2$O$_4$ is characteristic of the
spin-driven Jahn-Teller distortion that allows the system to order in a
N\'{e}el-type antiferromagnetic ground state.\cite{Kino1971} 
Dilute concentrations of Co result in this sharp drop being lost, which
indicates the structural distortion is likely suppressed and instead
the characteristic behavior is of spins freezing into a glassy state.  For
concentrations of Co above 60\%, N\'{e}el-type
ferrimagnetism is stabilized. The highest ordering temperatures  
$T_c$ for the different compounds are also listed in 
table\,\ref{table:magnetism}, obtained by taking the first derivative  
of the susceptibility with respect to temperature.
The frustration index $f = \Theta_{CW}/T_c$ is listed in the last column of the
table. It is seen that all the compounds order at temperatures much lower than
would be otherwise expected, \textit{ie.} all the compounds in the series
are frustrated. The initial increase in the frustration index with $x$ 
arises because of quenched disorder introduced by some of the A-sites having 
magnetic moments, and is not a true measure of geometric frustration.

\begin{figure}
\centering \includegraphics[width=9cm]{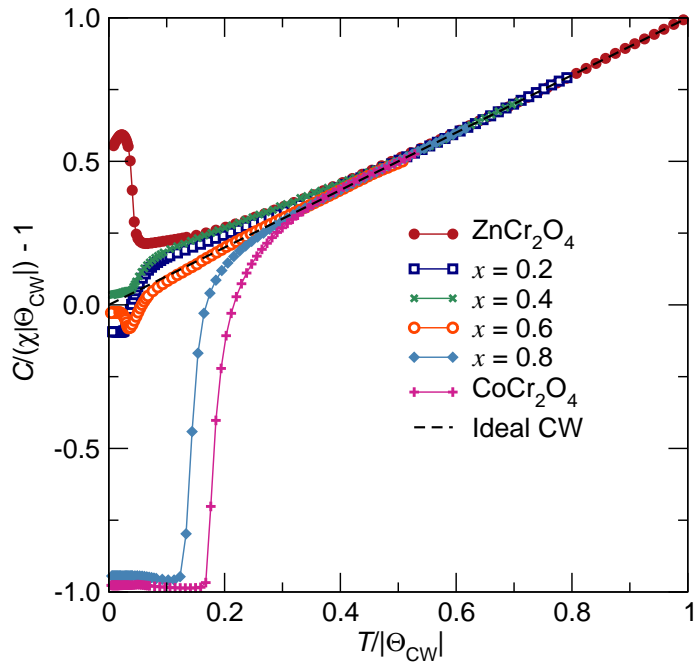}
\caption{(Color online) Scaled inverse susceptibility as a function of 
scaled temperature, as described by the formula in \ref{eqn:CW2}.
The dashed line represents ideal Curie-Weiss paramagnetism.
Positive deviations from this line reflect the presence of compensated 
antiferromagnetic interactions, while negative deviations reflect
uncompensated interactions and a tendency to ferrimagnetic ground states.
\label{fig:cw}}
\end{figure}

We can rearrange the Curie-Weiss equation: 

\begin{equation}
\chi  = \frac{C}{T - \Theta_{CW}} 
\label{eqn:CW1}
\end{equation}

\noindent in the following manner: 

\begin{equation}
\frac{C}{\chi \Theta_{CW}} = \frac{T}{\Theta_{CW}} - 1
\label{eqn:CW2}
\end{equation}

\noindent
A plot of $C/(\chi|\Theta_{CW}|) -1$ as a function of $T/|\Theta_{CW}|$ 
collapses all the high temperature susceptibility data 
in the manner displayed in figure\,\ref{fig:cw}. The straight line through
the origin corresponds to ideal Curie-Weiss behavior and can be used as a
measure of the quality of the fits in the high temperature regime. It is evident once
again that all magnetic ordering takes place at temperatures much smaller
than $T = \Theta_{CW}$ suggesting frustration.
(Note the horizontal axis in this plot is an inverse of the
frustration index $f$.) The modified Curie-Weiss plot displayed in 
figure\,\ref{fig:cw} also quickly allows one to distinguish the nature of the 
dominant magnetic exchange interactions. Positive deviations from the ideal 
Curie-Weiss line (dashed) correspond to samples with purely compensated 
antiferromagnetic interactions. 
Even small amounts of Co$^{2+}$ on the  A site result in uncompensated 
antiferromagnetism (ferrimagnetism) and this manifests as negative deviations
from the ideal line. As larger amounts of Co$^{2+}$ 
are substituted, this residual paramagnetism is lost and the ground states
resemble partially saturated ferrimagnets.

The modified Curie-Weiss plot depicted in figure\,\ref{fig:cw} superficially
resembles a semilog plot of resistivity \textit{versus} temperature for a
series of samples undergoing (typically) a composition driven metal-insulator 
transition. The line or curve separating insulators from metals is 
almost horizontal and frequently extrapolates, at 0\,K, to the inverse of the 
Mott minimum metallic conductivity.\cite{Mott1972,Mott1981} 
Insulators (localized systems) lie above this line, and metals (extended 
systems) lie below. In the same vein, samples in the series considered 
here show (appropriately scaled) inverse susceptibilities which lie above
the Curie-Weiss line provided all interactions are antiferromagnetic and
fully compensated, and these are distinct from samples with uncompensated
interactions or with ferromagnetic interactions which are analogues of 
extended states. Analogies with thermal expansion can be drawn as well. 
Localized (Einstein) modes can decrease thermal expansion coefficients in 
crystals, and even make it negative. Extended modes (Debye) usually give rise 
to the more common positive coefficients of thermal expansion.\cite{Ramirez1998}

\begin{figure}
\centering \includegraphics[width=12cm]{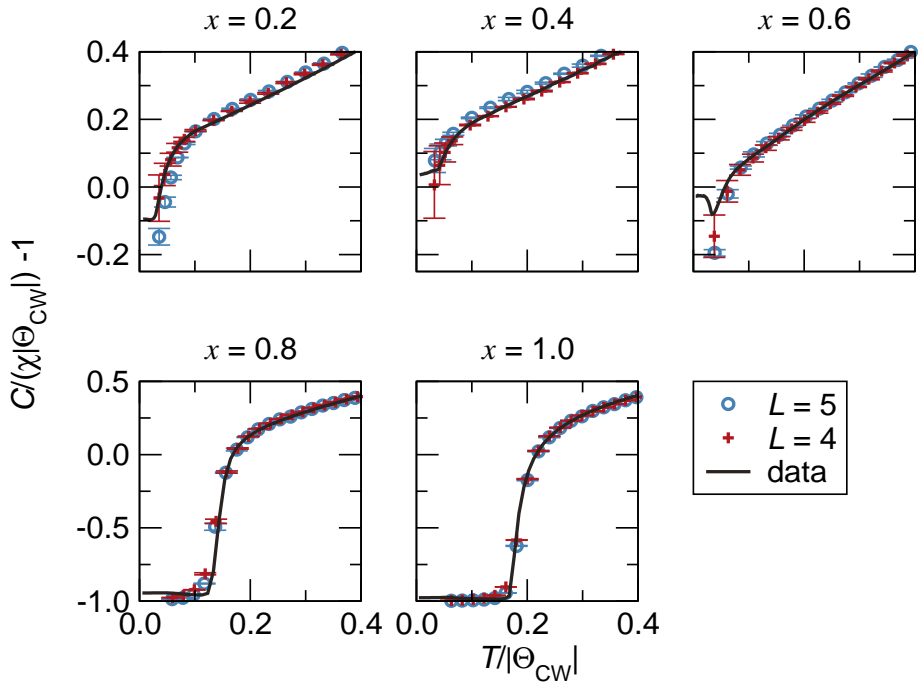}
\caption{Comparison of the experimental magnetic susceptibilities with 
Monte Carlo simulations performed for the corresponding $x$ values using the 
$J$ couplings listed in table\ref{table:best_fits}. $L$ corresponds to the
system size used in the simulation. 
\label{fig:simul1}}
\end{figure}

\begin{table}
\caption{Estimates of the Heisenberg coupling constants that best describe 
the experimental data. The $J_{BB}$ estimates were obtained by solving
equation~\ref{eqn:analytic_cw} for $J_{BB}$ using the 
$\Theta_{\mbox{\tiny CW}}$ estimates in table~\ref{table:magnetism}.
The range of $J_{BB}$ values for $x=0.2$ corresponds to a $J_{AA}/J_{BB}$ 
between 0.0 and 0.5. \label{table:best_fits}}
\begin{center}
\begin{tabular}{lcccccc}
\hline
\hline
\ & \multicolumn{2}{c}{$J_{AA}/J_{BB}$} & \multicolumn{2}{c}{$J_{AB}/J_{BB}$} & $J_{BB} (K)$ \\
     \          & estimate & error    & estimate    & error           & \   \\
\hline
ZnCr$_2$O$_4$   & -  & - & -  & - & -      \\
$x$ = 0.2       & \multicolumn{2}{c}{(no dependence)}  & 0.800 & 0.025   & 42.0-42.3  \\
$x$ = 0.4       & 0.3  & 0.1 & 0.775   & 0.025       & 41.5          \\
$x$ = 0.6       & 0.5  & 0.2 & 0.377   & 0.003       & 67.9          \\
$x$ = 0.8       & 0.4  & 0.1 & 0.930   & 0.02       & 35.4          \\
CoCr$_2$O$_4$   & 0.55  & 0.05 & 1.02   & 0.01       & 35.2         \\
\hline
\hline
\end{tabular}
\end{center}
\end{table}

It turns out that a simple quantitative understanding of the susceptibility 
data in figure~\ref{fig:cw} is possible as well.  We performed classical Monte
Carlo simulations  of the nearest-neighbor Heisenberg model on a B site spinel
with A site doping and found a series of Heisenberg coupling parameters
$J_{AA}/J_{BB}$ and $J_{AB}/J_{BB}$  that bring the simulations into rather
close agreement with the experimental susceptibility measurements.  

The best fits obtained are shown in figure~\ref{fig:simul1} and the
resulting coupling  constant estimates are summarized in
table~\ref{table:best_fits}.  The fits were first found using systems
consisting of $4^3$ conventional cubic unit cells  (1024 B sites) with periodic
boundary conditions. We then checked the finite-size scaling  on systems of
$5^3$ unit cells (2000 B sites). It should also be noted that these best fits 
were determined by searching the parameter space, not by an optimization
algorithm, so  that quite likely even better agreement is possible using the
same model. 

\begin{figure}
\centering \includegraphics[width=9cm]{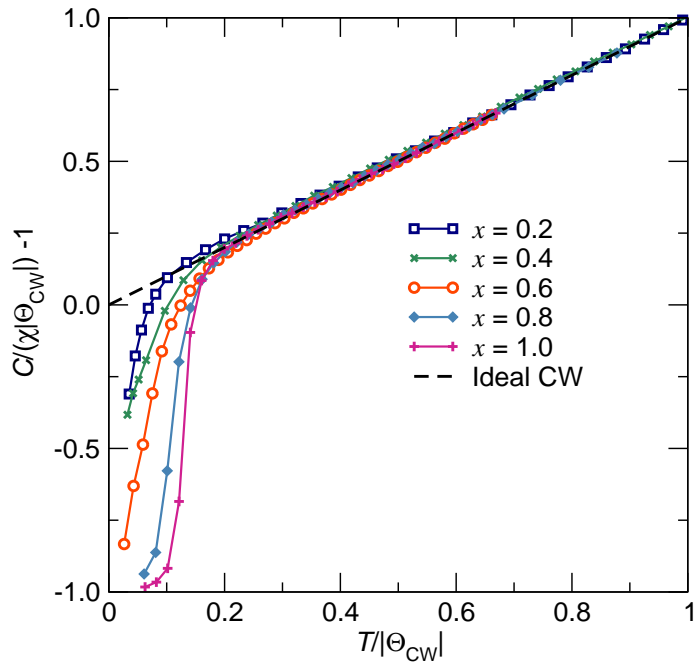}
\caption{Results of Monte Carlo simulations on Heisenberg spin systems
with a system size of $L = 4$ and the following coupling parameters:
$J_{AA}/J_{BB}$ = 0.45 and $J_{AB}/J_{BB}$ = 0.90.
\label{fig:simul2}}
\end{figure}

We were encouraged to see that all of the best $J_{AA}/J_{BB}$ and
$J_{AB}/J_{BB}$ parameters  do not differ much from each other (the $x=0.2$
data had too weak a dependence on $J_{AA}$ for us to determine it
effectively). Evidently,  the bulk magnetic behavior of the entire series of
materials is well described by taking  $J_{AB}$ to be comparable to $J_{BB}$
and taking $J_{AA}$ to be about half of $J_{BB}$. Accordingly, it was 
considered useful to actually fix all the couplings to such values  and only
vary $x$ -- the resulting scan (see figure~\ref{fig:simul2}) shows that it is indeed the value
of $x$ that most strongly  determines the bulk magnetic properties of the
system. The suprise is that for the range of parameters studied by which 
seem to be close to the experimentally measured susceptibility,
$J_{AB}$ and $J_{BB}$ are effectively equal, rather than $J_{AB}$ being
significantly larger than $J_{BB}$; the frequent expectation. Lyons, Kaplan, 
Dwight, and Menyuk (LKDM) \cite{Lyons1962} have proposed the following 
inequality: $(4J_{BB}S_B)/3J_{AB}S_A) \le 8/9$ implies collinear N\'eel ordering
of moments in spinels such as CoCr$_2$O$_4$. For $(4J_{BB}S_B)/3J_{AB}S_A) > 8/9$
non-collinear ordering is suggested. Since in CoCr$_2$O$_4$, $S_A$ = $S_B$,
we expect that if $J_{BB} \approx J_{AB}$, the ground state will be 
non-collinear, which is certainly true for CoCr$_2$O$_4$. So the Monte Carlo
estimates of $J_{BB}$ and $J_{AB}$ are consistent with the experimental 
realization of non-collinear ground states.\cite{Tomiyasu2004} 
Recently Ederer and Komeilj\cite{Ederer2007} have performed detailed
density functional calculations on the $x$ = 1 end-member, CoCr$_2$O$_4$.
They suggest that $J_{AA}$ is important (ignored by LKDM) as determined here
as well. While $J_{AA}$ in their calculations is determined to be typically 
less than $J_{BB}$, they estimate that $J_{AB}$ can be even twice as 
large as $J_{BB}$ which is not the result obtianed in the preliminary 
analysis presented here.

\begin{figure}
\centering \includegraphics[width=9cm]{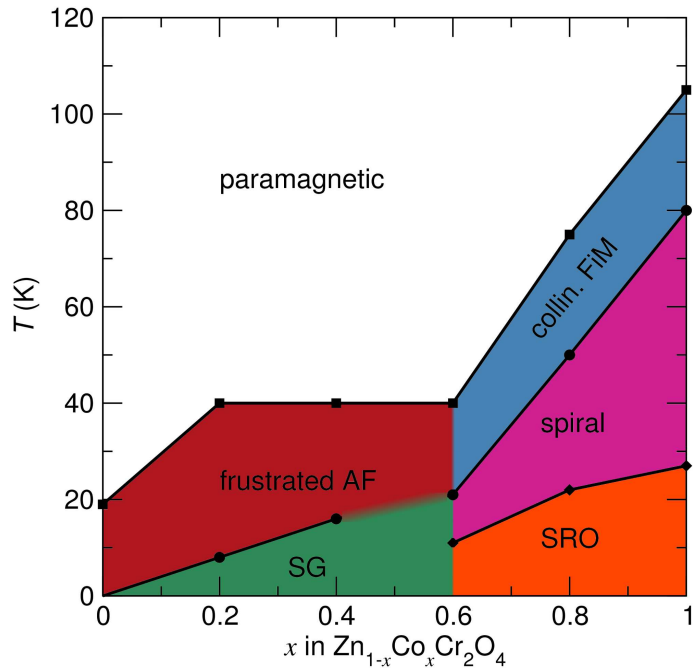}
\caption{Tentative magnetic phase diagram as a function of Co content
and temperature. Low concentrations of Co results in a glassy freezing of
the spins at low temperature while higher concentrations drive the system
to order in a non-collinear spiral structure. The points in the diagram
correspond to changes of slope in the susceptibility as a function of 
temperature. (SG = spin glass, AF = antiferromagnetism, FiM = ferrimagnetism,
SRO = short range ordering) 
\label{fig:phase}}
\end{figure}

The results of the field-cooled susceptibility traces was used to construct a 
tentative magnetic phase diagram for the solid solutions as depicted in 
figure\,\ref{fig:phase}. We use changes in the slope of the field-cooled
susceptibility as a function of temperature to obtain the boundaries between
different magnetic phases. Regardless of composition, all samples demonstrate 
paramagnetic behavior at high temperatures and begin to show transitions to 
long range order starting near 100\,K. The end member ZnCr$_2$O$_4$ is the 
only compound which exhibits long range antiferromagnetic order because of
its  structural distortion. We make the reasonable assumption that even small
amounts of Co result in glassy behavior as indicated by the line separating
frustrated antiferromagnetism from glassy states. For concentrations of  $x
\ge$ 0.6 the magnetic behavior begins to look more like that of  CoCr$_2$O$_4$
with a collinear ferrimagnetic transition between 50\,K and  100\,K followed
by a transition into a non-collinear spiral at lower  temperatures.  
The behavior near $x$ = 0.6 is rather complex and more experiments will
be particularly important for this region of the phase diagram. The
transition temperatures for the spiral appear to be  slightly suppressed by Zn
substitution with respect to the end member as  illustrated by the positive
slope of the phase boundary as the CoCr$_2$O$_4$  end member is approached.
For the Co-rich compounds at the lowest temperatures, the structures begin to
develop some spiral short range order (SRO).\cite{Tomiyasu2004}

In summary, we have prepared and studied the magnetic properties of a solid 
solution of ZnCr$_2$O$_4$ and CoCr$_2$O$_4$.    We have found that replacing
non-magnetic cations with magnetic cations in the tetrahedral sites of a well
ordered spinel results in a transition from glassy behavior at low
concentrations of magnetic cations into a N$\acute{e}$el-like ferrimagnetic
order as well as commensurate and incommensurate magnetic spirals.  These
results are significant because they demonstrate that the presence of magnetic
cations on the A-site can have the effect of relieving magnetic frustration on
the B-site by allowing the system to find a lower energy state in the form of
non-collinear magnetic spirals.  Monte Carlo simulations suggest the minimum
physiccs required to describe trends in the magnetic behavior, and suggest the
range of strengths of the different magnetic couplings. We also present a 
useful form for plotting susceptibility data as a function of temperature which
allows us to easily recognize the existence of uncompensated magnetic 
interactions in otherwise antiferromagnetic systems. 

\section{Acknowledgements} EMS would like to acknowledge encouraging  
discussions with Leon Balents. The National Science
Foundation for support through a Career Award (NSF-DMR\,0449354), and
for the use of MRSEC facilities (Award NSF-DMR\,0520415). JED was supported 
by the RISE program at the UCSB MRL.

\clearpage


\medskip

\begin{thebibliography}{10}

\bibitem{Ramirez1994}
Ramirez A~P 1994 {\em Annu. Rev. Mater. Sci.\/} {\bf 24} 453

\bibitem{KimuraAnnuRev2007}
Kimura T 2007 {\em Annu. Rev. Mater. Res.\/} {\bf 37} 387

\bibitem{Kimura2003}
Kimura T, Goto T, Shintani H, Ishizaka K, Arima T and Tokura Y 2003 {\em
  Nature\/} {\bf 426} 55

\bibitem{Katsura2005}
Katsura H, Nagaosa N and Balatsky A~V 2005 {\em Phys. Rev. Lett.\/} {\bf 95}
  057205

\bibitem{Kenzelmann2005}
Kenzelmann M, Harris A~B, Jonas S, Broholm C, Schefer J, Kim S~B, Zhang C~L,
  Cheong S~W, Vajk O~P and Lynn J~W 2005 {\em Phys. Rev Lett.\/} {\bf 95}
  087206

\bibitem{Mostovoy2006}
Mostovoy M 2006 {\em Phys. Rev. Lett.\/} {\bf 96} 067601

\bibitem{Sergienko2006}
Sergienko I~A and Dagotto E 2006 {\em Phys. Rev. B\/} {\bf 73} 094434

\bibitem{Yamasaki2006}
Yamasaki Y, Miyasaka S, Kaneko Y, He J-P, Arima T and Tokura Y 2006
{\em Phys. Rev. Lett.} {\bf 96} 207204

\bibitem{Lawes2006}
Lawes G, Melot B, Page K, Ederer C, Hayward M~A, Proffen T and Seshadri R 2006
{\em Phys. Rev. B\/} {\bf 74} 024413

\bibitem{Martinho2001}
Martinho H, Moreno N O, Sanjurjo J A, Rettori C, Garc\'ia-Adeva A J, Huber D L, Oseroff S B, Ratcliff W, Cheong S.-W, Pagliuso P G, Sarrao J L, and Martins G B 2001 {\em Phys. Rev. B\/} {\bf 64} 024408

\bibitem{Sianou2005}
Sianou A K, Stergioudis G A, Efthimiadis K G, Kalogirou O, and Tsoukalas I A 2005 {\em J. Alloys Compd.\/} {\bf 392} 310

\bibitem{Cheong2007}
Cheong S~W and Mostovoy M 2007 {\em Nature Mater.\/} {\bf 6} 13

\bibitem{Tackett2007}
Tackett R, Lawes G, Melot B~C, Grossman M, Toberer E~S and Seshadri R 2007 {\em
  Phy. Rev. B\/} {\bf 76} 024409

\bibitem{Fritsch2004}
Fritsch V, Hemberger J, B\"uttgen N, Scheidt E~W, Krug~von Nidda H~A, Loidl A
  and Tsurkan V 2004 {\em Phys. Rev. Lett.\/} {\bf 92} 116401

\bibitem{Buettgen2004}
B\"{u}ttgen N, Hemberger J, Fritsch V, Krimmel A, M\"{u}cksch M, von Nidda
  H~A~K, Lunkenheimer P, Fichtl R, Tsurkan V and Loidl A 2004 {\em New J.
  Phys.\/} {\bf 6} 191

\bibitem{Bergman2007}
Bergman D, Alicea J, Gull E, Trebst S and Balents L 2007 {\em Nature Phys.\/}
  {\bf 3} 487

\bibitem{Blasse1963}
Blasse G 1963 {\em Philips Res. Rept.\/} {\bf 18} 383

\bibitem{Baltzer1965}
Baltzer P~K, Lehmann H~W and Robbins M 1965 {\em Phys. Rev. Lett.\/} {\bf 15} 
493

\bibitem{Baltzer1966}
Baltzer P~K, Wojtowicz P~J, Robbins M and Lopatin E 1966 {\em Phys. Rev.\/}
 {\bf 151} 367

\bibitem{Ramirez1997}
Ramirez A~P, Cava R~J and Krajewski J 1997 {\em Nature\/} {\bf 386}
 156

\bibitem{Rudolf2007}
Rudolf T, Kant C, Mayr F, Hemberger J, Tsurkan V and Loidl A 2007 {\em New J.
  Phys.\/} {\bf 9} 76

\bibitem{Ortega-San-Martin2008}
Ortega-San-Mart\'{i}n L, Williams A~J, Gordon C~D, Klemme S and Attfield J~P
  2008 {\em J. Phys. Cond. Matter\/} {\bf 20} 104238 

\bibitem{Tokura2008}
Ohgushi K, Okimoto Y, Ogasawara T, Miyasaka S and Tokura Y 2008 {\em J. Phys.
  Soc. Jpn.\/} {\bf 77} 034713

\bibitem{Kino1971}
Kino Y and L\"{u}thi B 1971 {\em Solid State Comm.\/} {\bf 9} 805

\bibitem{Lee2000}
Lee S~H, Broholm C, Kim T~H, Ratcliff W and Cheong S~W 2000 {\em Phys. Rev.
  Lett.\/} {\bf 84} 3718

\bibitem{Tomiyasu2004}
Tomiyasu K, Fukunaga J and Suzuki H 2004 {\em Phys. Rev. B\/} {\bf 70}
  214434

\bibitem{Berar1998}
B\'erar J and Baldinozzi G 1998 {\em IUCr-CPD Newsletter\/} {\bf 20} 3

\bibitem{ALPS}
 Albuquerque A~F et al. 2007  {\em J.\ Magn.\ Magn.\ Mater.\/} {\bf 310} 1187

\bibitem{Shannon1976}
Shannon R~D 1976 {\em Acta Crystallogr. A\/} {\bf 32} 751

\bibitem{Hill1979}
Hill R~J, Craig J~R and Gibbs G~V 1979 {\em Phys. Chem. Minerals\/} {\bf 4}
  317

\bibitem{Nakatsuka2006}
Nakatsuka A, Ikeda Y, Nakayama N and Mizota T 2006 {\em Acta Crystallogr. E\/}
  {\bf 62}(5) i109

\bibitem{Cossee_JPhysChemSolids1960}
Cossee P and van Arkel A~E 1960 {\em J. Phys. Chem. Solids\/} {\bf 15} 1

\bibitem{Yan2008}
Yan L~Q, Maci\'{a} F, Jiang Z~W, Shen J, He L~H and Wang F~W 2008 {\em J. Phys.
  Condensed. Matter\/} {\bf 20} 255203

\bibitem{Blazey1966}
Blazey K~W 1966 {\em Solid State Comm.\/} {\bf 4} 541

\bibitem{Mott1972}
Mott N~F 1972 {\em Philos. Mag.\/} {\bf 26} 1015

\bibitem{Mott1981}
Mott N~F 1981 {\em Philos. Mag. B\/} {\bf 44} 265

\bibitem{Ramirez1998}
Ramirez A~P and Kowach G~R 1998 {\em Phys. Rev. Lett.\/} {\bf 80}
  4903--4906

\bibitem{Lyons1962}
Lyons D~H, Kaplan T~A, Dwight, K and Menyuk N 1962
{\em Phys Rev\/} {\bf 126} 540

\bibitem{Ederer2007}
Ederer C and Komelj M 2007
{\em Phys Rev B\/} {\bf 76} 064409

\bibitem{Tomiyasu2006}
Tomiyasu K and Itoh S 2006 {\em J. Phys. Soc. Jpn.\/} {\bf 75} 084708

\end{thebibliography}
\end{document}